\documentclass[10pt,preprint]{aastex}

\shorttitle{Obtaining Potential Field Solution}
\shortauthors{T\'oth et al.}

\begin{document}

\title{Obtaining Potential Field Solution with Spherical Harmonics 
       and Finite Differences}

\author{G\'abor T\'oth, Bart van der Holst, and Zhenguang Huang}
\affil{Center for Space Environment Modeling,
       University of Michigan, Ann Arbor, MI 48109}

\newcommand{\bB}{\mathbf{B}}

\begin{abstract}
Potential magnetic field solutions can be obtained 
based on the synoptic magnetograms of the Sun.
Traditionally, a spherical harmonics decomposition of the magnetogram 
is used to construct the current and divergence free magnetic field
solution. This method works reasonably well when the order of spherical 
harmonics is limited to be small relative to the resolution of the magnetogram,
although some artifacts, such as ringing, can arise around sharp features.
When the number of spherical harmonics is increased, however, 
using the raw magnetogram data given on a grid that is uniform
in the sine of the latitude coordinate 
can result in inaccurate and unreliable results, especially
in the polar regions close to the Sun.

We discuss here two approaches that can mitigate or completely 
avoid these problems: 
i) Remeshing the magnetogram onto a grid with uniform 
resolution in latitude, and limiting the highest order of the
spherical harmonics to the anti-alias limit;
ii) Using an iterative finite difference algorithm to solve for 
the potential field.
The naive and the improved numerical solutions are compared
for actual magnetograms, and the differences are found to be
rather dramatic. 

We made our new Finite Difference Iterative Potential-field Solver
(FDIPS) a publically available code, so that other researchers can
also use it as an alternative to the spherical harmonics approach.
\end{abstract}


\section{Introduction}

Magnetograms provide the radial magnetic field on the visible surface of the
Sun. The actual measurement is for the line-of-sight component of the magnetic field,
which is then transformed into the radial component assuming an (approximately)
radial field near the solar surface.
As the Sun rotates, the individual magnetograms can be combined into a
synoptic magnetogram that covers the whole spherical surface.
Synoptic magnetograms are provided by many observatories,
including Wilcox Solar Observatory (WSO), the 
Michelson Doppler Imager (MDI) instrument on the Solar and Heliospheric
Observatory (SOHO), the Global Oscillation Network Group (GONG),
Solar Dynamic Observatory (SDO) and the 
Synoptic Optical Long-term Investigations of the Sun (SOLIS) observatory.
Today's magnetograms contain hundreds to thousands of pixels along each 
coordinate direction. 
These magnetograms can be used to extrapolate the magnetic
field into the solar corona.

The simplest model \citep{Schatten:1969}
assumes a current-free, in other words potential, 
magnetic field that matches the radial field of
the magnetogram on the surface, while it satisfies a simple
boundary condition at the outer boundary at some radial distance $R$.
The outer boundary condition is usually taken at $R=2.5\,$R$_s$
(solar radii), and a purely radial field is assumed at this ``source
surface''.  Mathematically the problem is the following: given the
magnetogram data that defines the radial component of the magnetic field 
as $M(\theta,\phi)$ at $r=1\,$R$_s$, find the scalar potential $\Phi$ so that
\begin{eqnarray}
  \nabla\cdot(\nabla \Phi)              &=& 0     \label{eq:laplace} \\ 
  \left.\frac{\partial \Phi}
             {\partial r}\right|_{r=1}  &=& M(\theta,\phi)  
                                                  \label{eq:innerbc} \\
  \left.\Phi\right|_{r=R}               &=& 0     \label{eq:outerbc}
\end{eqnarray}
Here $\theta \in [0,\pi]$ and $\phi \in [0,2\pi]$ are the co-latitude and 
longitude coordinates, repectively. 
Once the solution is found, the potential field solution is obtained as
\begin{equation}
 \bB = \nabla \Phi
\end{equation}
and it will trivially satisfy both the divergence-free and the 
current-free properties
\begin{eqnarray}
  \nabla \cdot  \bB = \nabla\cdot(\nabla \Phi)    &=& 0     \\
  \nabla \times \bB = \nabla \times (\nabla \Phi) &=& 0    
\end{eqnarray}
We note that the current is only zero inside the domain. If the
solution is continued out to $r>R$ with a purely radial magnetic field
$B_r(r>R) = (R/r)^2 B_r(R)$, there will be a finite current at $r\ge R$,
on the other hand, the divergence will be zero for all $r>1$.

The potential field solution is often obtained with a spherical
harmonics expansion \citep{Altschuler:1977}. Here we briefly
summarize the procedure in its simplest possible form.
The base functions
$\varphi_{nm}$ are the spherical harmonic functions 
$Y_{nm}$ multiplied with an appropriate linear combination
of the corresponding radial functions $r^n$ and $r^{-n-1}$ so 
that the boundary condition $\varphi_{nm}(R,\theta,\phi)=0$ 
is satisfied:
\begin{equation}
  \varphi_{nm}(r,\theta,\phi) = 
        (r^n - R^{2n+1}r^{-n-1}) Y_{nm}(\theta,\phi)
\label{eq:phi}
\end{equation}
The indexes $n\ge0$ and $m$ ($|m|\le n$) are the integer degree and order 
of the spherical harmonic function, respectively. 
The $\varphi_{nm}$ functions are
solutions of the Laplace equation (\ref{eq:laplace}), satisfy the
boundary condition at $r=R$, and they form an orthogonal base in the 
$\theta,\phi$ coordinates. 
The magnetic potential solution can be approximated as a
linear combination of the base functions
\begin{equation}
  \Phi(r,\theta,\phi) = \sum_{n=1}^N \sum_{m=-n}^{n} f_{nm} 
        \varphi_{nm}(r,\theta,\phi)
\label{eq:Phi}
\end{equation}
where $N$ is the highest degree considered in the expansion and the
$n=m=0$ harmonics is not included, as it corresponds to the monopole term. 
The coefficients $f_{nm}$ can be determined by taking the radial derivative 
of equation (\ref{eq:Phi}) and equating it with the magnetogram radial
field at $r=1$:
\begin{equation}
  M(\theta,\phi) = \sum_{n=1}^N \sum_{m=-n}^{n} f_{nm} 
        \left.\frac{\partial \varphi_{nm}}{\partial r}\right|_{r=1}
    = \sum_{n=1}^N [n + (n+1)R^{2n+1}] 
      \sum_{m=-n}^{n} f_{nm} Y_{nm}(\theta,\phi)
\label{eq:M}
\end{equation}
Exploiting the orthogonality of the base functions,
we can take the inner product with $Y_{nm}$ to determine $f_{nm}$ as
\begin{equation}
  f_{nm} = \frac{1}{4\pi[n + (n+1)R^{2n+1}]}
       \int_0^\pi d\theta \sin \theta \int_0^{2\pi} d\phi 
       M(\theta,\phi) Y_{nm}(\theta,\phi)
\label{eq:integral}
\end{equation}
where the $1/(4\pi)$ coefficient results 
from the normalization of the spherical harmonics. 
An alternative approach of obtaining
the harmonic coefficients $f_{nm}$ is to employ 
a (least-squares) fitting procedure in equation (\ref{eq:M}). 
This is much more expensive than evaluating the integral in (\ref{eq:integral}), 
but it can be more robust if the magnetogram does not cover (well) the whole
surface of the Sun. 

Using the spherical harmonic coefficients the potential 
can be determined on an arbitrary grid using (\ref{eq:Phi})
and the magnetic field can be obtained with finite differences.
Alternatively, one can calculate the gradient of the base functions 
analytically and obtain the magnetic field as
\begin{equation}
  \bB(r,\theta,\phi) = \sum_{n=1}^N \sum_{m=-n}^{n} f_{nm} 
        \nabla \varphi_{nm}(r,\theta,\phi)
\label{eq:Bpot}
\end{equation}
for $1\le r \le R$.
Spherical harmonics provide a computationally efficient and very elegant
way of solving the Laplace equation on a spherical shell. However,
one needs to be cautious of how the integral in equation 
(\ref{eq:integral}) is evaluated, especially when a large number
of harmonics are used in the series expansion.

\begin{figure}
\plotone{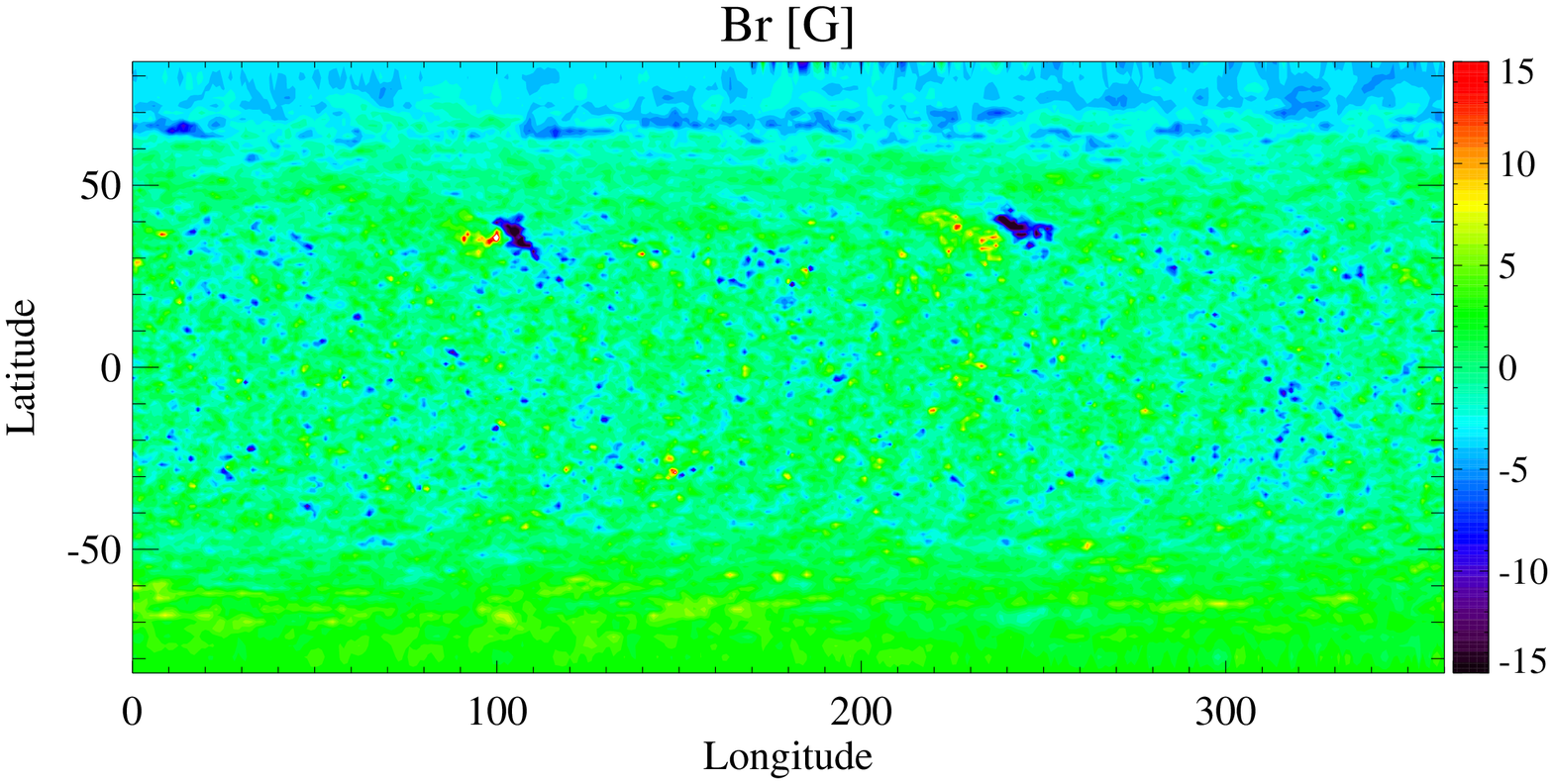}
\caption{Synoptic magnetogram for Carrington Rotation 2077 obtained by GONG. 
         The radial magnetic field at the photosphere is shown in the range
         $-15$ to $+15$\,Gauss to show more detail. 
         The color scale is saturated in the active regions where
         the largest and smallest values are $-45.4\,$G and $27.9\,$G.}
\label{fig:magnetogram}
\end{figure}

We will use the GONG synoptic magnetogram for Carrington Rotation 2077 
(CR2077, from November 20 to December 17, 2008)
as an example to demonstrate the problem. The magnetogram contains 
the radial field on a $180\times 360$ latitude-longitude grid on the 
solar surface.
The grid spacing is uniform in $\cos\theta$ (or sine of the latitude)
and in longitude $\phi$. Figure~\ref{fig:magnetogram} shows the
radial field. 

Section 2 discusses the naive and more sophisticated ways of  
obtaining the potential field solution with spherical harmonics.
Section 3 describes an alternative approach using an iterative 
finite difference. The various methods are compared in the final
Section 4, where we also demonstrate the ringing effect that can
arise in the spherical harmonics solution, and we draw our conclusions.

\section{Potential Field Solution Based on Spherical Harmonics}

To turn the analytic prescription given in the introduction into
a scheme that works with real magnetograms, one has to pick 
the maximum degree $N$, 
and evaluate the integrals in equation (\ref{eq:integral}) 
for each pair of $n$ and $m$ up to the highest order.
The resulting $f_{nm}$ coefficients can be used to construct the 3D 
potential magnetic field solution at any given point using 
equation (\ref{eq:Bpot}). 

\subsection{Naive Spherical Harmonics Approach}

The simplest approximation to equation (\ref{eq:integral}) 
is a discrete integral using the original magnetogram data:
\begin{equation}
  f_{nm} = \frac{1}{4\pi[n + (n+1)R^{2n+1}]}
       \sum_{i=1}^{N_{\theta}} \sum_{j=1}^{N_{\phi}}  
       (\Delta\cos\theta)_i (\Delta\phi)_j
       M_{i,j} Y_{nm}(\theta_i,\phi_j)
\label{eq:sum}
\end{equation}
where the $M_{i,j}$ is the radial field in a pixel of the 
$N_{\theta}$ by $N_{\phi}$ sized magnetogram. The pixel is
centered at the $(\theta_i,\phi_j)$ coordinates, and the area 
of the pixel is given by $(\Delta\cos\theta)_i(\Delta\phi)_j$.

\begin{figure}
\plotone{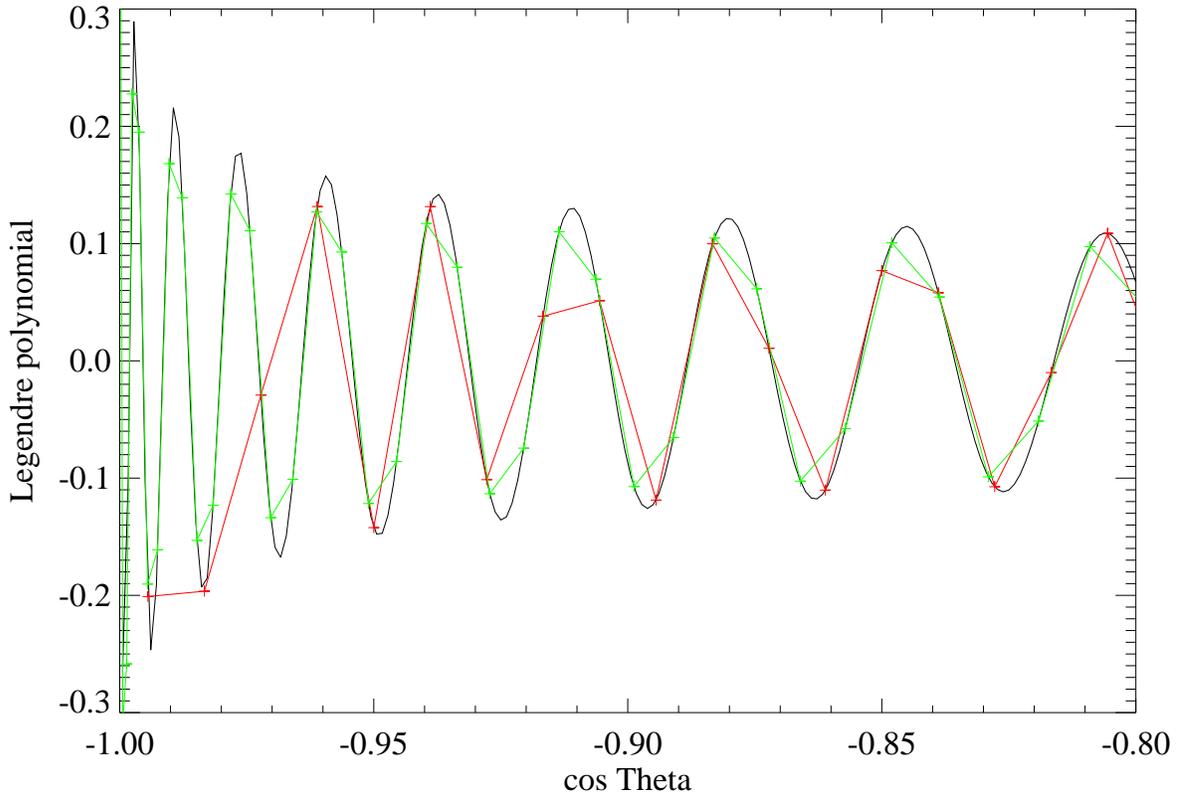}
\caption{Discrete representations of the $P_{90,0}$ associated Legendre 
         polynomial as a function of $\cos\theta$ near the ``south pole'' 
         at $\cos\theta=-1$. 
         The black curve shows an accurate representation with 1800 grid 
         points uniformly distributed in the $[-1,1]$ range. 
         The red curve represents the polynomial on 180 grid points uniform
         in $\cos\theta$, while the green curve uses 181 grid points that
         are uniformly distributed in $\theta$ including the poles at
         $\theta=0$ and $\theta=\pi$.}
\label{fig:legendre}
\end{figure}

Unfortunately, the uniform $\cos\theta$ mesh used by most of the magnetograms
is not at all optimal to evaluate the integral in equation (\ref{eq:integral}).
In fact this procedure will only work with maximum order $N$ that is 
much less than $N_{\theta}$.
Figure~\ref{fig:legendre} shows the $P_{90,0}$ associated Legendre
polynomial discretized in different ways. The red curve shows the
discretization on 180 grid points that are uniform in $\cos\theta$.
Clearly the red curve is a very poor representation near the poles,
where $\cos\theta = \pm 1$. This is important, because the 
amplitude of the Legendre polynomial
is actually largest near the poles. This means that the orthonormal
property is not satisfied in the discrete sense, and the coefficients
obtained with equation (\ref{eq:sum}) are very inaccurate. 
The Legendre polynomial can be represented much better on a uniform $\theta$
grid (shown by the green curve in Figure~\ref{fig:legendre}), 
as we will discuss below.

\begin{figure}
\plotone{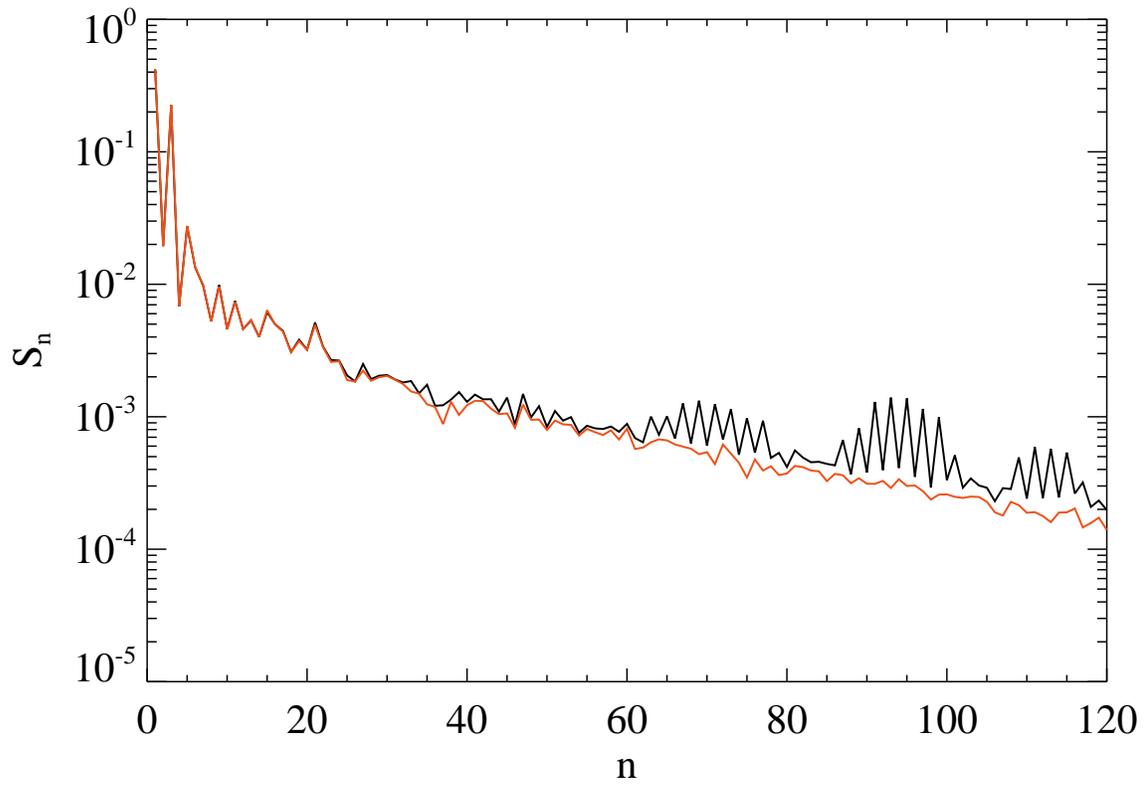}
\caption{Power spectrum $S_n = \sum_m f^2_{nm}$ of the spherical
         harmonics expansion with the original (black line) and remeshed
         (red line) magnetograms.}
\label{fig:sn}
\end{figure}

A clear signal of this problem is that the amplitudes of the higher order
spherical harmonics are not getting smaller with increasing indexes
$n$ and $m$, i.e., the harmonic expansion is not converging. The black line
in Figure~\ref{fig:sn} shows the amplitudes $S_n = \sum_m f^2_{nm}$ 
which is oscillating wildly for $n>60$ for this $360\times 180$ magnetogram.
The oscillations are almost exclusively due to the 
$f_{n0}$ coefficients, the $m\ne0$ coefficients are well behaved.
This plot can be directly compared with Figure 15 in \cite{Altschuler:1977},
where the power spectrum is more-or-less exponentially decaying.
We believe that the reason is that these authors used a least-square fitting 
to the line-of-sight (LOS) magnetic field instead of calculating the spherical
harmonics from the radial field as shown above. While the two methods
are identical analytically (assuming that the LOS and radial fields
correspond to the same solution), the use of least-square fitting 
mitigates the lack of orthogonality among the discretized Legendre funtions,
while the naive approach described above heavily relies on the
orthogonality property.

\begin{figure}
\plotone{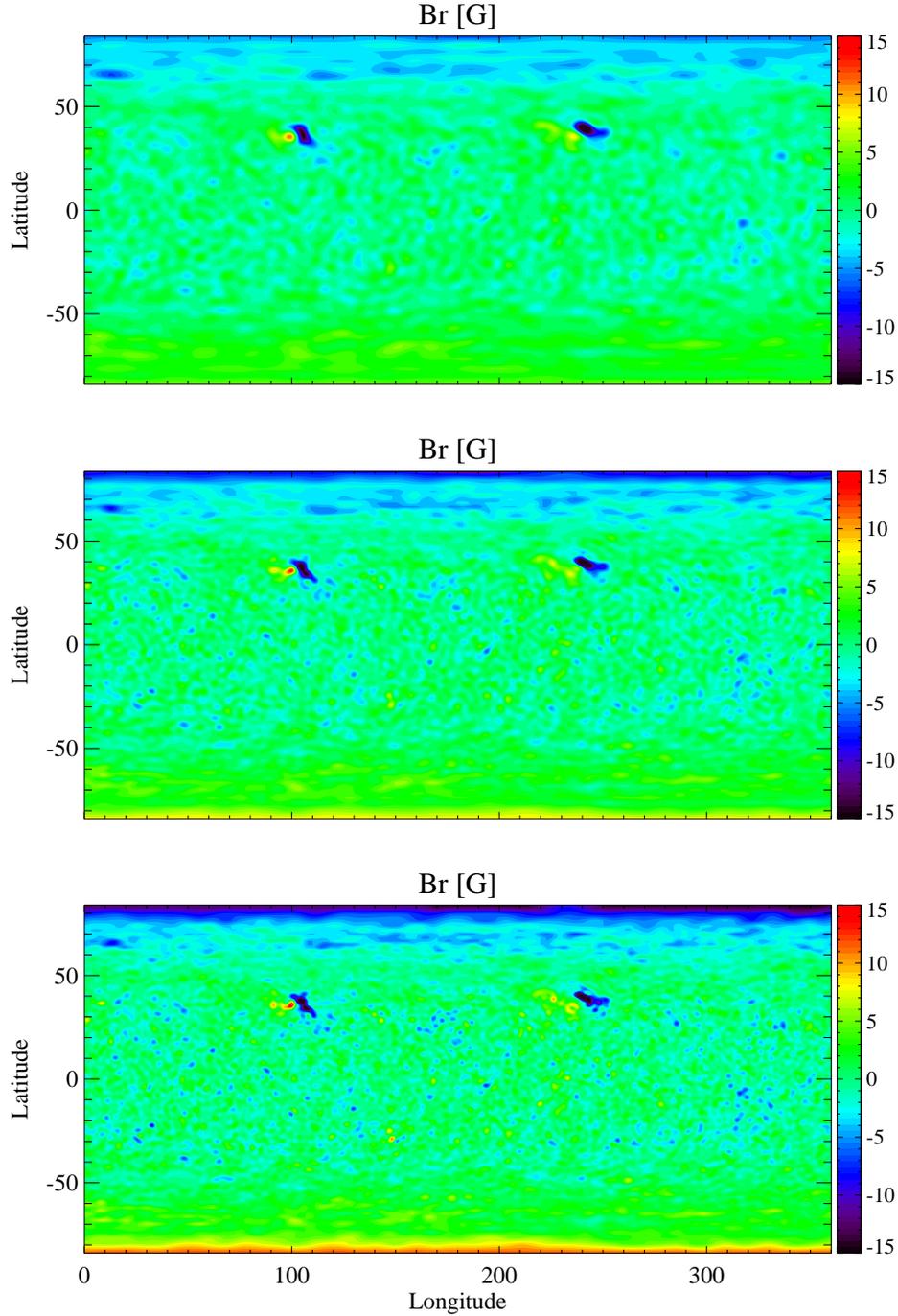}
\caption{Comparison of the radial magnetic field at $r=1$ using the
         spherical harmonics expansion on the original magnetogram grid
         up to $N=60$ (top), $N=90$ (middle)
         and $N=120$ (bottom) order. These plots should be compared
         with the magnetogram shown in Figure~\ref{fig:magnetogram}.
         Note that the magnetic field in the polar regions is completely
         wrong for $N=90$ and $N=120$. The $N=60$ solution is reasonable,
         but the fine details are smoothed out.}
\label{fig:naive_fft}
\end{figure}

Given the non-converging series expansion, 
the resulting potential field will be very inaccurate in the polar regions
and will have essentially random values depending on the number of 
spherical harmonics used. This is demonstrated by Figure~\ref{fig:naive_fft}
that shows the radial magnetic field reconstructed with various
number of harmonics using the original magnetogram grid. One would expect the
radial component of the potential field to reproduce the magnetogram shown in
Figure~\ref{fig:magnetogram}. Instead, we find that the solution deviates
strongly in the polar region if the harmonics expansion is continued above
$N=60$. For $N=60$ (or lower) the solution looks reasonable, but strongly
smoothed due to the insufficient number of harmonics. 
This is most obvious around
the active regions in the top panel of Figure~\ref{fig:naive_fft}.

We note that these numerical errors are not related or comparable to the 
observational uncertainties of the magnetograms, which are usually also
quite large in the polar regions. The observational uncertainties
are essentially unavoidable but are within some well-understood
range. On the other hand, these numerical artifacts are definitely avoidable
while the errors are essentialy unbounded if one uses too many harmonics.

\subsection{Spherical Harmonics with Remeshed Magnetogram}

One can get much more accurate results if the magnetogram is remeshed 
to a grid that is uniform in the co-latitude $\theta$, has an odd number
of nodes, and contains the two poles $\theta=0$ and $\theta=\pi$. 
In fact this is the standard grid used in spherical harmonics transforms 
(e.g. \cite{Suda:2001}) and it is often referred to as using the 
Chebyshev nodes, since the uniform $\theta$ grid points correspond 
to the Chebyshev nodes in the original $\cos\theta$ coordinate which 
is the argument of the Legendre polynomials.
Figure~\ref{fig:legendre} shows that the Legendre polynomial is much better
represented on the uniform $\theta$ grid than on the uniform $\cos\theta$ grid.
Remeshing the magnetogram introduces some new adjustable parameters
into the procedure: the number of grid cells $N'_{\theta}$
on the new mesh, and the interpolation procedure.

If the remeshing is done with the same number of grid points as is in
the original magnetogram grid, the latitudinal cell size at the equator
will be a factor of $\pi/2$ larger than in the uniform-$\cos\theta$ grid. 
On the other hand, the uniform-$\theta$ grid will contain
many more points than the original in the polar regions,
so the interpolation procedure may create some unwanted artifacts. 
To maintain the resolution of the original data around the equator, we set
$N'_{\theta}$ to $(\pi/2)N_{\theta}$ rounded to an odd integer.
For the remeshing we chose a simple linear interpolation procedure, 
and it works satisfactorily, but one could certainly use higher order 
interpolation procedures, such as splines. 
Before doing the interpolation, we add extra grid cells
corresponding to the north and south poles of the magnetogram grid,
and the values at these two extra cells are set as the average of the pixels
around the poles:
\begin{eqnarray}
M_{0} &=& 
       \frac{1}{N_{\phi}}\sum_{j=1}^{N_{\phi}} M_{1,j} \\
M_{N_{\theta}+1} &=& 
       \frac{1}{N_{\phi}}\sum_{j=1}^{N_{\phi}} M_{N_{\theta},j}
\end{eqnarray}
The co-latitude coordinates of the uniform-$\theta$ mesh are
\begin{equation}
  \theta'_{i'}=\pi \frac{i'-1}{N'_{\theta}-1}
\end{equation}
for $i'=1\ldots N'_{\theta}$. We use simple linear interpolation from the 
extended magnetogram mesh to the uniform $\theta$ mesh:
\begin{equation}
M'_{i',j} = \alpha M_{i,j} + (1-\alpha) M_{i'+1,j}
\end{equation}
where the $i$ index is determined so that 
$\theta_i \le \theta'_{i'} \le \theta_{i+1}$ and 
\begin{equation}
   \alpha = \frac{\theta_{i+1} - \theta'_{i'}}
                 {\theta_{i+1} - \theta_i}
\end{equation}
Finally the spherical harmonics coefficients are determined with the 
integral approximated as
\begin{equation}
  f_{nm} = \frac{1}{4\pi[n + (n+1)R^{2n+1}]}
       \sum_{i=1}^{N'_{\theta}} \sum_{j=1}^{N_{\phi}}  
       \epsilon_i w_i(\Delta\phi)_j
       M'_{i,j} Y_{nm}(\theta'_i,\phi_j)
\label{eq:sum2}
\end{equation}
where $\epsilon_1=\epsilon_{N'_{\theta}}=1/2$ and $\epsilon_i=1$ for all
other indexes. The $w_i$ coefficients are the Clenshaw-Curtis weights 
\citep{Clenshaw:1960, Potts:1998} defined as
\begin{equation}
  w_i = \frac{1}{H} \sum_{k=0}^{H} \epsilon'_k 
             \frac{-2}{4k^2-1}\cos(2\pi k\theta'_i)
\end{equation}
where $H=(N'_{\theta}-1)/2$ and $\epsilon'_0=\epsilon'_H=1/2$ and 
$\epsilon'_k=1$ for all other indexes. We note that for $N'_{\theta}$
order of 10 or more
\begin{equation}
w_i \approx \frac{|\cos\theta'_{i_+} - \cos\theta'_{i_-}|}{2}
\end{equation}
where  $i_+ = \min(i+1,N'_{\theta})$ and $i_- = \max(i-1,1)$ are the
indexes of the neighboring cells, or the cell itself at the poles.

\begin{figure}
\plotone{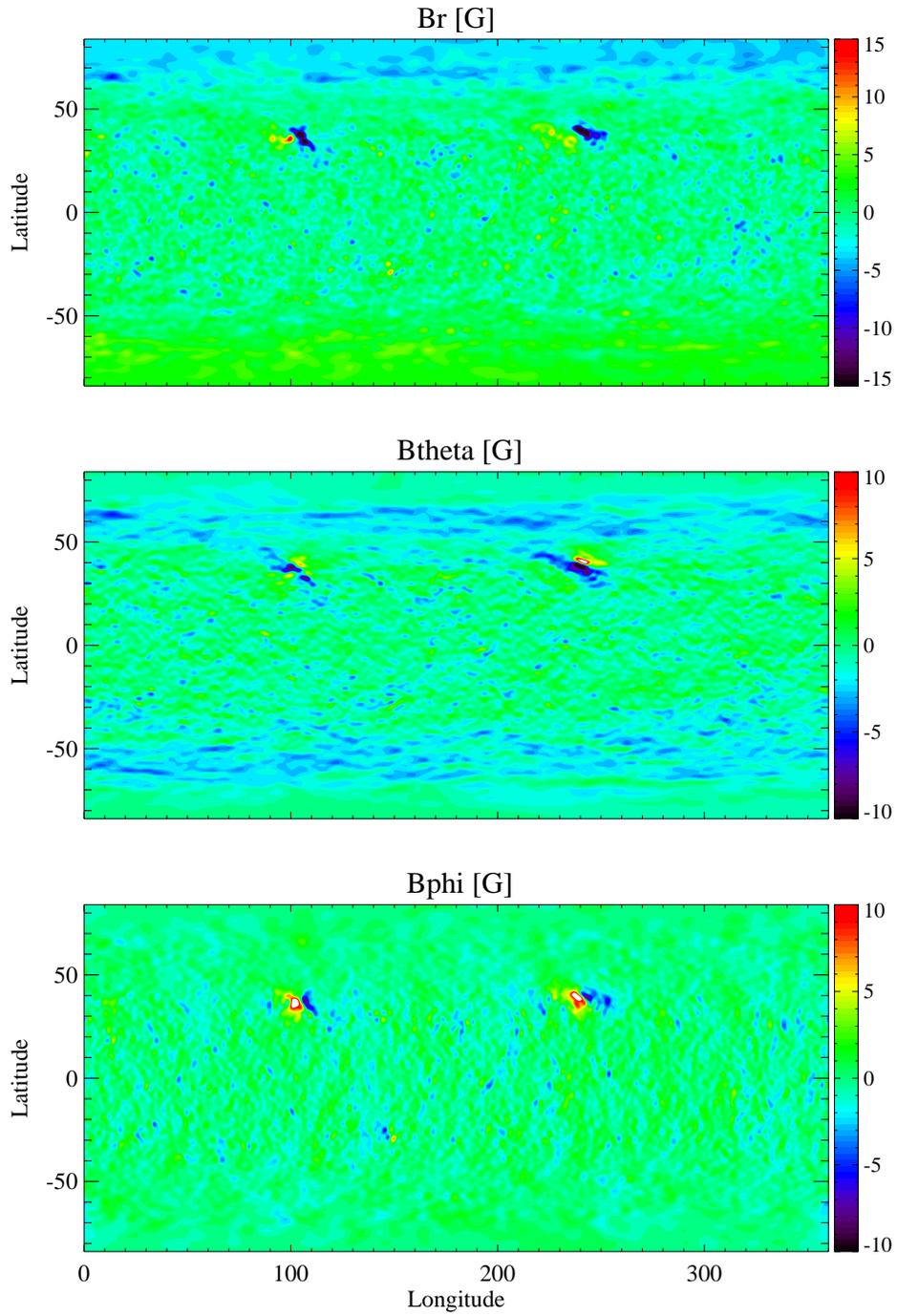}
\caption{Magnetic field solution at $r=1$ using the remeshed
         spherical harmonics up to $N=120$ degree.
         The radial compnents (top panel) well reproduces the 
         magnetogram shown in Figure~\ref{fig:magnetogram},
         although some of the details are slightly smoothed out.}
\label{fig:remeshed_fft}
\end{figure}

Using the proper grid allows us to use larger number of spherical harmonics.
$N$ is limited only by the $N \le 2N'_{\theta}/3$ and $N \le N_{\phi}/3$
alias-free conditions \citep{Suda:2001}. For our example 
$180\times 360$ magnetogram, we remesh it a to $283\times360$ uniform-$\theta$
grid, and we can obtain accurate solutions up to $N=120$ degree 
spherical harmonics.
Figure~\ref{fig:remeshed_fft} shows the potential field solution obtained with 
the remeshed magnetogram grid with $N=120$. 
Compared with the naive method, the solution is much more reasonable
in the polar regions. There is some smoothing when compared to the magnetogram
shown in Figure~\ref{fig:magnetogram}, most obvious near the active regions.

While the remeshing is definitely a big improvement over using the
original magnetogram, it would be nice to be able to use the original
magnetogram data without remeshing and interpolation. The next
section shows that this can be easily achieved with a finite
difference solver.

\section{Finite Difference Iterative Potential Field Solver (FDIPS)}

The Laplace equation (\ref{eq:laplace}) with the boundary conditions 
(\ref{eq:innerbc}) and (\ref{eq:outerbc}) can be solved quite easily
with an iterative finite difference method. The advantage of 
finite differences compared with spherical harmonics is that
the boundary data given by the magnetogram directly effects the solution
only locally, while the spherical harmonics are global functions,
and their amplitudes depend on all of the magnetogram data. If the
magnetogram contains large discontinuities, we expect the finite difference
scheme to be better behaved. 

The finite difference method has advantages if the solution is to be
used in a finite difference code on the same grid, because one can guarantee
zero divergence and curl for the magnetic field in the finite difference sense.
The solution obtained with the spherical harmonics has zero
divergence and curl analytically, but not on the finite difference grid,
which may severly underresolve the high order spherical harmonic functions in
some regions (see Figure~\ref{fig:legendre}).

The finite difference method was applied to the solar potential magnetic field 
problem as early as 1976 (\cite{Adams:1976}), but the method was limited
by the computational resources available at the time. 
Solving a 3D Laplace equation on today's computers is an almost trivial 
problem. We implemented the new Finite Difference Iterative Potential-field Solver
(FDIPS) code in Fortran 90. The serial version does 
not require any external libraries, while the parallel version uses the 
Message Passing Interface (MPI) library for communication.
FDIPS can solve the Laplace equation on a $150\times180\times360$ spherical grid 
to high accuracy on a single processor in less than an hour. 
The parallel code can solve the same problem in less than 5 minutes on 16 processors.

We briefly describe the algorithm in FDIPS. We use a staggered spherical grid: 
the magnetic field is discretized on cell faces while the potential is 
discretized at the cell centers. We use one layer of ghost
cells to apply the boundary conditions so the cell centers are
located at $r_i,\theta_j,\phi_k$ with $i=0\ldots N_r+1$, 
$j=0\ldots N_{\theta}+1$ and 
$k=0\ldots N_{\phi}+1$. 
The $\theta_j$ and $\phi_k$ coordinates
of the real cells are given by the magnetogram, while the ghost cell
coordinates are given by $\theta_0 = -\theta_1$, 
$\theta_{N_{\theta}+1} = 2\pi - \theta_{N_{\theta}}$,
$\phi_0 = \phi_{N_{\phi}}$, and $\phi_{N_{\phi}+1} = \phi_1$.
We allow for a non-uniform radial grid extending from $r=1$ to $R$, but
for sake of simplicity in this paper a uniform radial grid is used with
$r_i = 1+(i-1/2)\Delta r$ with $\Delta r = (R-r)/N_r$. 

The radial magnetic field components are located at the radial cell interfaces
at $(r_{i+1/2},\theta_j,\phi_k)$ where $r_{i+1/2}=(r_i+r_{i+1})/2$ for 
$i=0\ldots N_r$, $j=1\ldots N_{\theta}$ and $k=1\ldots N_{\phi}$. 
Similarly the latitudinal components are at 
$r_i,\theta_{j+1/2},\phi_k$ with 
$\cos\theta_{j+1/2}=(\cos\theta_j+\cos\theta_{j+1})/2$, and 
$j=0\ldots N_{\theta}$. Note that the interface is taken half-way in the
$\cos\theta$ coordinate and not in $\theta$, because this makes the cells equal 
area when the magnetogram is given on uniform $\cos\theta$ grid.
Finally the longitudinal field components are located at
$(r_i,\theta_j,\phi_{k+1/2})$ where $\phi_{k+1/2}=(\phi_k+\phi_{k+1})/2$
for $k=0\ldots N_{\phi}$.

The staggered discretization keeps the stencil of the Laplace operator compact 
and it makes the boundary condtions relatively simple. 
The magnetic field is obtained as a discrete gradient of $\Phi$:
\begin{eqnarray}
  B_{r,i+1/2,j,k}      &=& \frac{\Phi_{i+1,j,k}-\Phi_{i,j,k}}
                                {\Delta r}\nonumber\\
  B_{\theta,i,j+1/2,k} &=& 
                        \frac{\sin\theta_{j+1/2}(\Phi_{i,j+1,k}-\Phi_{i,j,k})}
                             {r_i \Delta\cos\theta} \nonumber\\
  B_{\phi,i,j,k+1/2}   &=& \frac{\Phi_{i,j,k+1}-\Phi_{i,j,k}}
                                {r_i \sin\theta_j \Delta\phi}
\label{eq:gradient}
\end{eqnarray}
Note the $\sin\theta/\Delta\cos\theta$ factor in
the $\theta$ derivative for the uniform $\cos\theta$ grid.
For uniform $\theta$ grid this is replaced with $1/\Delta\theta$.
The divergence of the magnetic field, i.e. the Laplace of $\Phi$
is obtained as
\begin{eqnarray}
  0 = (\nabla^2 \Phi)_{i,j,k} &=& 
      \frac{r^2_{i+1/2}B_{r,i+1/2,j,k} - r^2_{i-1/2,j,k}B_{r,i-1/2,j,k} }
           {r^2_i \Delta r}                                 \nonumber\\
    &+&\frac{\sin\theta_{j+1/2}B_{\theta,i,j+1/2,k} 
          - \sin\theta_{j-1/2}B_{\theta,i,j-1/2,k}    }
           {r_i \Delta\cos\theta}                           \nonumber\\
    &+&\frac{ B_{\phi,i,j,k+1/2} -  B_{\phi,i,j,k-1/2} }
           {r_i\sin\theta_j\Delta\phi }
\label{eq:laplacenum}
\end{eqnarray}
Again $1/\Delta\cos\theta$ is used for the uniform $\cos\theta$ grid, 
while on the uniform $\theta$ grid
this is replaced with $1/(\sin\theta\Delta\theta)$.

The magnetogram boundary condition is applied by setting the inner ghost cell as
\begin{equation}
 \Phi_{0,j,k} = \Phi_{1,j,k} - \Delta r\, M'_{j,k}
\label{eq:innerbcnum}
\end{equation}
where $M'_{j,k} = M_{j,k}-\bar M$ is the magnetogram with the average field
(i.e. the monopole due to observation errors)
\begin{equation}
  \bar M = \frac{1}{4\pi}\sum_{j,k} (\Delta\cos\theta)_j (\Delta\phi)_k M_{j,k}
\end{equation}
removed.
The zero potential at $r_{N_r+1/2}=R$ is enforced by setting the ghost cell as
\begin{equation}
  \Phi_{N_r+1,j,k} = -\Phi_{N_r,j,k}
\label{eq:outerbcnum}
\end{equation}
The boundary conditions at the poles are a bit tricky. Cells $(i,1,k)$ and
$(i,1,k')$ are on opposite sides of the north pole if 
$k'=\textrm{mod}(k-1+N_{\phi}/2, N_{\phi})+1$. Therefore the ghost cells
in the $\theta$ direction are set as $\Phi_{i,0,k}=\Phi_{i,1,k'}$ and
$\Phi_{i,N_{\theta}+1,k}=\Phi_{i,N_{\theta},k'}$. We note here that $N_{\phi}$
is assumed to be an even number.
The periodic boundaries in the $\phi$ direction are simple:
$\Phi_{i,j,0}=\Phi_{i,j,N_{\phi}}$ and 
$\Phi_{i,j,N_{\phi}+1}=\Phi_{i,j,1}$.

We need to find $\Phi_{i,j,k}$ that satisfies the discrete Laplace
equation (\ref{eq:laplacenum}) with the boundary conditions applied
via the ghost cells. The initial guess is $\Phi=0$ which provides a non-zero
residual because of the inhomogeneous boundary condition at the inner
boundary applied by equation (\ref{eq:innerbcnum}). 
We use this residual with a negative sign as the righ-hand-side of 
the Poisson equation $(\nabla^2\Phi)_{i,j,k}=R_{i,j,k}$, 
and use $\Phi_{0,j,k}=\Phi_{1,j,k}$ as the new
homogeneous inner boundary condition instead of 
(\ref{eq:innerbcnum}). We use the Krylov-type iterative 
method BiCGSTAB \citep{vanderVorst:1992} to find the solution.
The linear system is preconditioned with an Incomplete Lower-Upper decomposition 
(ILU) preconditioner to speed up the convergence. We use ILU(0) with no
fill-in compared to the original matrix structure, so the preconditioner
is a diagonal matrix, but its elements depend on all elements of the original
matrix. We have implemented a serial as well as a 
parallel version of the algorithm. In the parallel version the
preconditioner is applied separately in each subdomain. 
FDIPS finds an accurate (down to $10^{-10}$ relative error) solution 
on a $180^3$ grid in less than 1000 iterations. 
Even running serially, this takes less than an hour on today's computers.

Once the solution is found in terms of the discrete potential $\Phi_{i,j,k}$, 
we apply the original boundary conditions including (\ref{eq:innerbcnum})
and calculate the magnetic field with equation(\ref{eq:gradient}). 
The divergence 
of the magnetic field will be zero to the accuracy of the Poisson solver.
The curl of the magnetic field will be zero in a finite difference sense
simply because it is constructed as the discrete gradient of the potential.
The boundary condition at the inner boundary is also satisfied exactly:
$B_{r,i=1/2,j,k}=M'_{j,k}$. Averaging
$r B_{\phi}$ and $r B_{\theta}$ to the $r_{N_r+1/2} = R$ position
at the outer boundary also gives exactly zero tangential fields due to the
equations (\ref{eq:outerbcnum}) and (\ref{eq:gradient}).
Depending on the application, we may interpolate the potential magnetic field 
onto a collocated grid, or use it on the original staggered grid.

\begin{figure}
\plotone{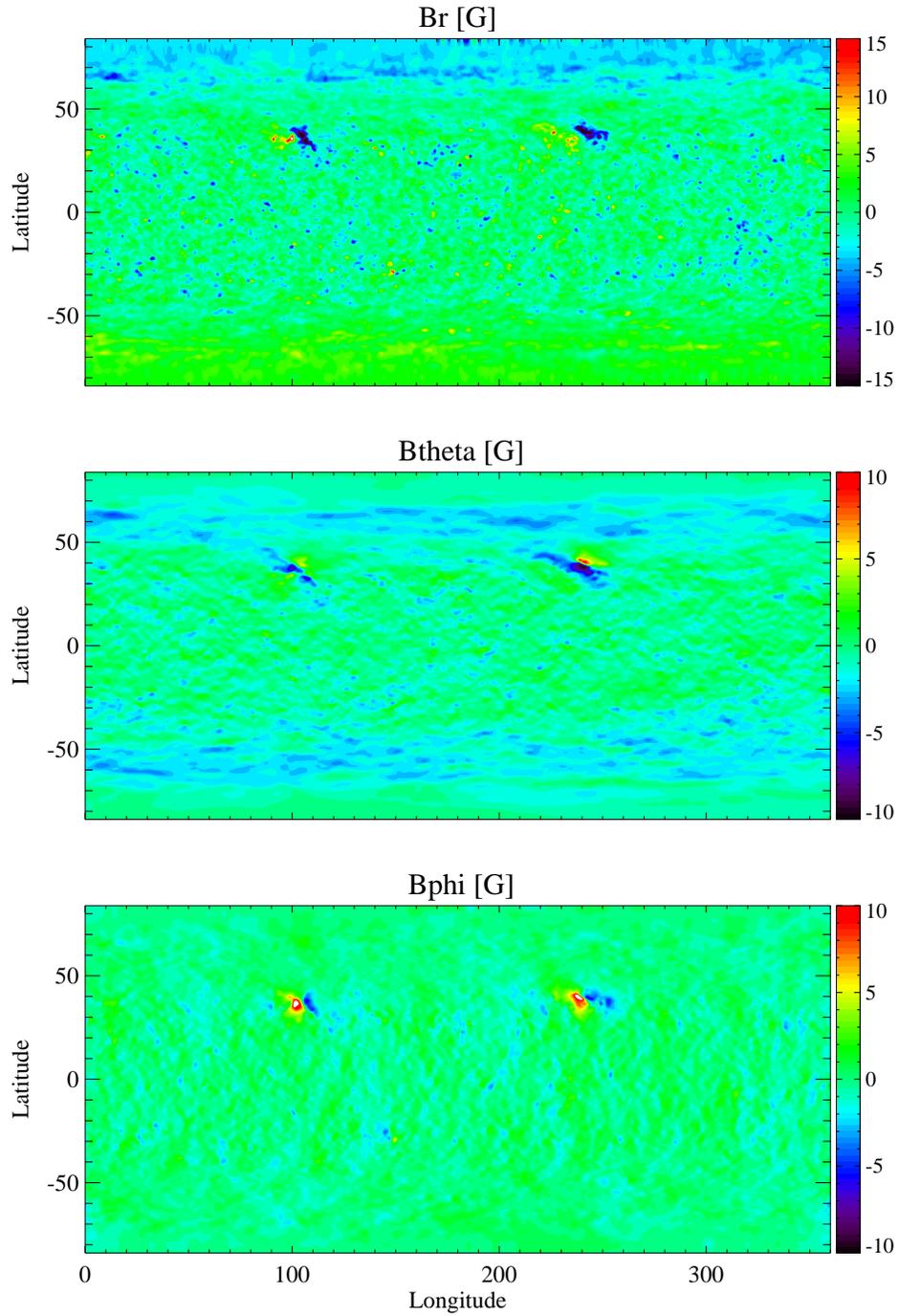}
\caption{Magnetic field solution at $r=1$ using the finite difference 
         code FDIPS on a $150\times 360\times 180$ grid.
         The radial component (top panel) agrees exactly with the
         magnetogram shown in Figure~\ref{fig:magnetogram} except
         for the removal of the average field (the monopole).
         The tangential components (middle and bottom panels) 
         agree well with the remeshed 
         spherical harmonics solution shown in Figure~\ref{fig:remeshed_fft}.
	 }
\label{fig:fdips}
\end{figure}

Figure~\ref{fig:fdips} shows the solution of the magnetic field obtained 
with the finite difference solver FDIPS on a $150\times 360\times 180$ grid. 
Since we use the same uniform-$\cos\theta$ grid as the magnetogram, 
the obtained radial field is identical with the magnetogram at $r=1$.
The tangential components agree well with the remeshed spherical harmonics 
solution shown in Figure~\ref{fig:remeshed_fft}. It took 1166 iterations
to get a solution with a relative accuracy of $10^{-10}$. The run time was
almost exactly one hour on a 2.66 GHz Intel CPU. 

\section{Discussion and Conclusions}

We have discussed various ways to obtain the potential field solution
based on solar magnetograms. While spherical harmonics provide an 
efficient and elegant method, there are some subtle restrictions that
require attention. If one wants to use many spherical harmonics
(the same order as the number of magnetogram pixels in the colatitude 
direction), the magnetogram data on the $N_{\theta}\times N_{\phi}$ grid 
has to be remeshed 
onto a uniform-$\theta$ grid with $N'_{\theta}\times N_{\phi}$ points, 
$N'_{\theta}$ must be an odd number, and the new grid must include both poles.
After the remeshing the maximum degree of harmonics $N$ is only
limited by the anti-alias limit to $\min(2N'_{\theta}/3$, $N_{\phi}/3)$.
We used a simple linear interpolation for the remeshing.

The remeshing can be avoided by the use of a 3D finite difference scheme.
One can use the original magnetogram grid, and the only freedom is in
choosing the radial discretization. The finite difference scheme provides
a solution that is fully compatible with the boundary conditions,
and the solution has zero divergence and curl in the finite difference sense.

\begin{figure}
\plottwo{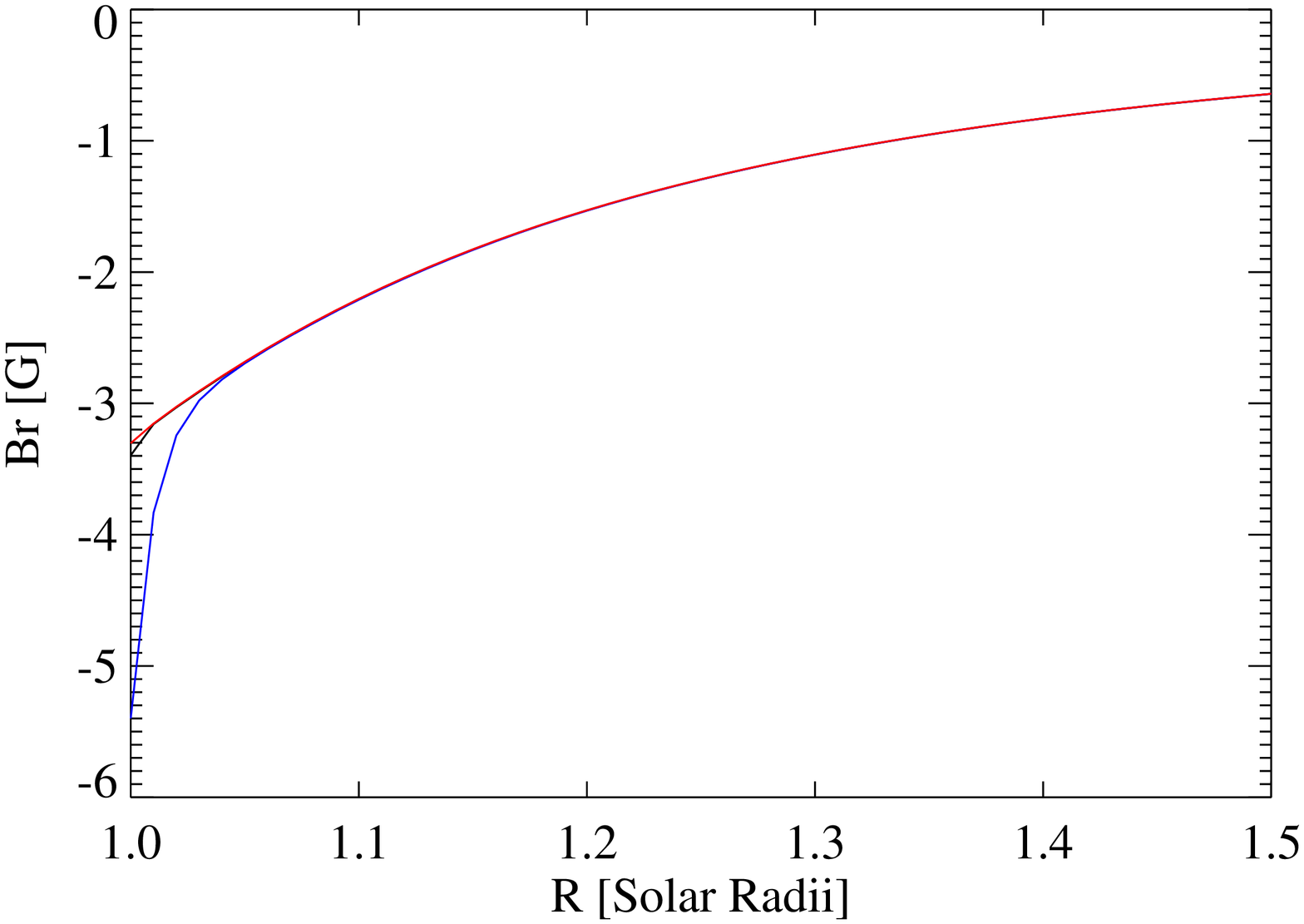}{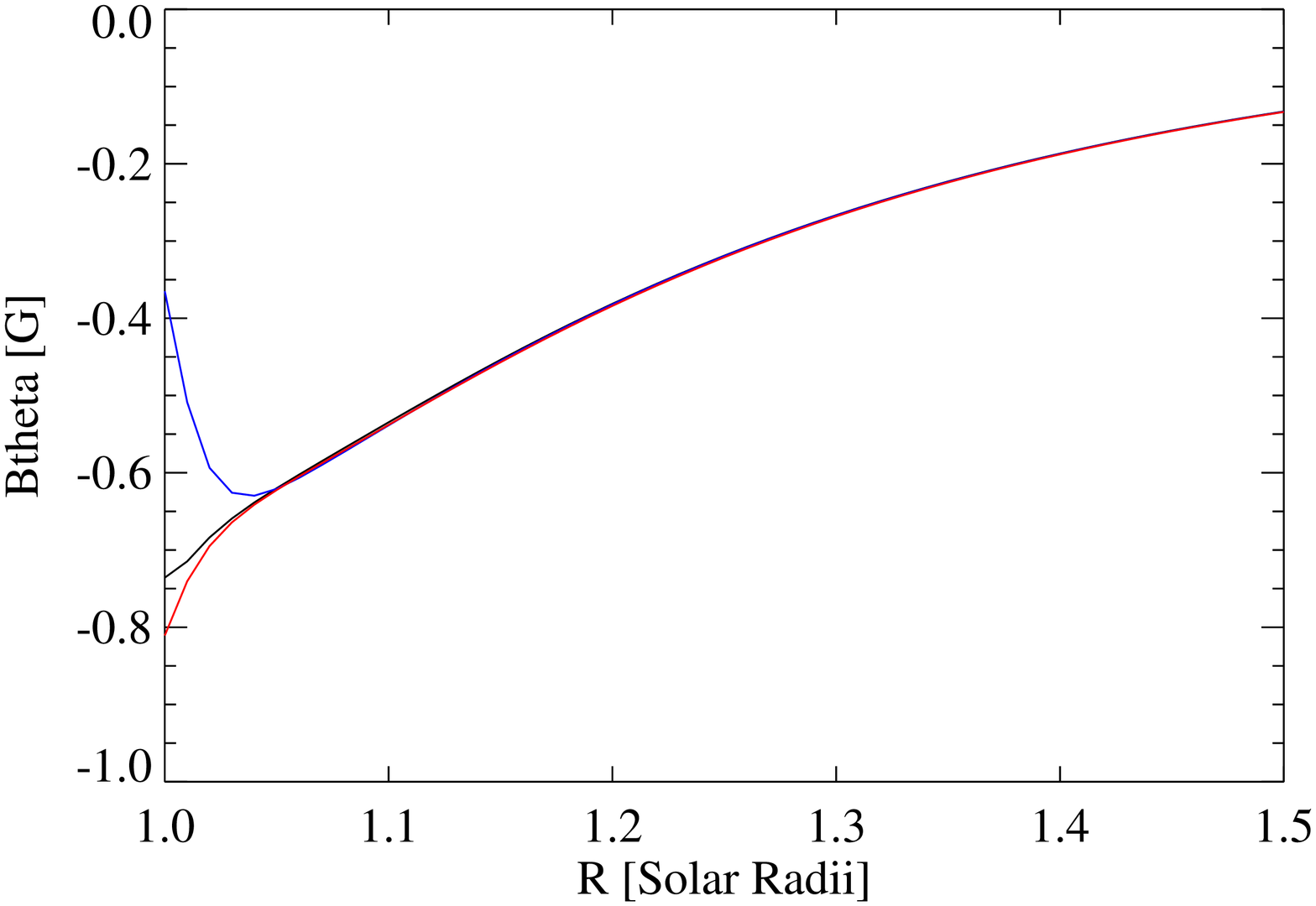}
\caption{The radial and latitudinal components of the magnetic
         field along the radial coordinate at fixed $76.5\degr$ latitude
         and $30\degr$ longitude. The solutions are obtained with the 
         naive (blue line) and remeshed (red line) harmonics approaches
         with $N=120$, as well as with finite differences (black line).
}
\label{fig:radial}
\end{figure}

Figure~\ref{fig:radial} compares the solutions obtained with the three methods 
along the radial direction for a fixed latitude $\theta=76.5\degr$ and
longitude $\phi=30\degr$. The spherical harmonics series were truncated 
at $N=120$ for both the naive and remeshed methods.
The naive spherical harmonics algorithm gives incorrect 
results close to the solar surface where the high order harmonics dominate.
This is most obvious for the radial component, which is given at $r=1$
by the magnetogram, and it is exactly reproduced by the finite difference 
scheme. The latitudinal component at $r=1$ is also very different from 
the values given by the remeshed harmonics and the finite differences. 
The latter two methods agree reasonably well with each other.
For radial distances above $r=1.05$ all three methods agree quite well.

So far we restricted our example to a GONG magnetogram taken 
at the solar minimum. If one uses an MDI magnetogram during solar maximum,
the largest magnetic fields are much stronger (order of 1000\,G) and
the resolution of the magnetogram is much finer (order of 1000 pixels). 
The finer magnetogram resolution allows going to larger number of harmonics,
even when using the original magnetogram grid (naive approach).
But the strong and sharp gradients in the magnetogram will bring out another
problem with the spherical harmonics approach, the ringing effect. 
The ringing is due to the so-called Gibbs phenomenon: the step-function like
magnetogram data results in high amplitude high order harmonics in 
Fourier space. The ringing effect and other
artifacts are discussed in great detail by \cite{Tran:2009}.

\begin{figure}
\plotone{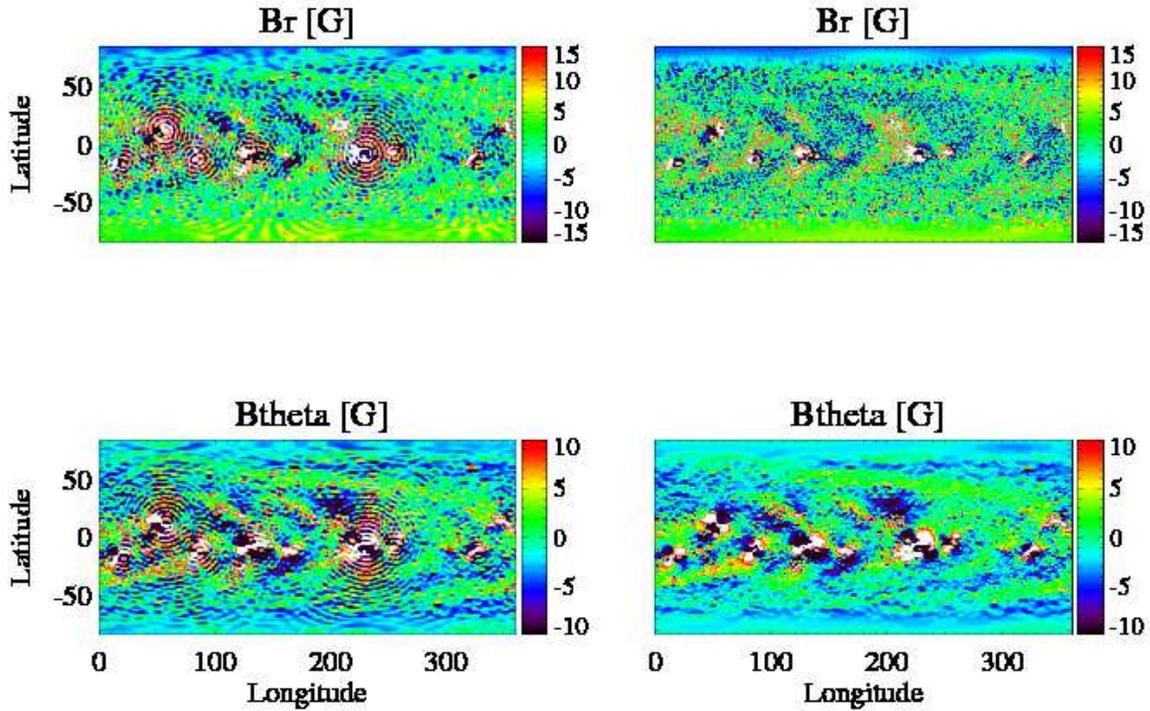}
\caption{Radial and latitudinal components of the potential field solution 
         at $r=1$ for the MDI magnetogram for CR2029
         obtained with remeshed spherical harmonics algorithm with
         maximum order $N=90$ (left) and with FDIPS using a 
         $150\times 360 \times 180$ grid (right). The color scale is
         saturated in the active regions.
}
\label{fig:ringing}
\end{figure}

Figure~\ref{fig:ringing} demonstrates this effect on
the $3600\times1080$ resolution MDI magnetogram for Carrington Rotation 2029
(from April 21 to May 18, 2005),
with the maximum radial field strength around $\pm3000\,$G. 
The remeshed harmonics method with $N=90$ is compared with the finite
difference method on a $150\times 360\times 180$ grid (the magnetogram
data is coarsened to a $360\times 180$ grid). In the spherical 
harmonics solution the ringing is very clearly visible around the 
active regions, both in the radial and latitudinal components.
The finite difference scheme, on the other hand, shows no sign of
ringing in either components. This is obvious for the radial component,
which simply coincides with the coarsened magnetogram, but for the
latitudinal component it is due to the fact that
the finite difference solution of the Laplace equation 
does not suffer from ringing artifacts even for discontinuous boundary data. 
For the spherical harmonics approach the ringing
becomes weaker with increased number of harmonics, but it is quite
apparent even for $N=180$ (not shown). The results of the remeshed and
naive harmonics methods are essentially the same up to $N=180$,
i.e. the ringing is not due to the remeshing of the magnetogram. 

In terms of computational efficiency, a good implementation of the
spherical harmonics scheme is much faster than the finite difference scheme.
In fact, it may be more costly to construct the potential field 
solution on a 3D grid from the spherical harmonics coefficients 
than obtaining the coefficients themselves. Our Fortran 90 code can
obtain the spherical coefficients up to $N=60$, 90 and 120 degrees 
in 1, 1.8 and 3.3 seconds, respectively, while the
reconstruction of the solution on the $151\times 361\times 180$
grid takes 5, 12, and 20 minutes, respectively. All timings were
done on a single 2.66 GHz Intel CPU.
The reconstruction cost can be improved by running the code in parallel,
and/or truncating the series in parts of the grid where the higher
order harmonics have a negligible contribution. We also note that
going beyond about $N=360$ harmonics becomes fairly complicated 
\citep{Potts:1998}. 

The computational cost of the finite difference scheme scales 
with the number of grid cells and the number of iterations.
The number of iterations is fairly constant for multigrid
type methods, but for the Krylov sub-space schemes it grows 
with the problem size, although slower than linearly.
The finite difference scheme can be sped up by parallelizing the
code, which is fairly straightforward for the Krylov subspace schemes.
Since we limit the ILU preconditioning to operate independently on 
the subdomains of each processor, the preconditioner
becomes less efficient as the number of processors increases,
which results in an increase in the number of iterations.
To minimize this effect, the parallel FDIPS code splits the grid in the 
$\theta$ and $\phi$ directions only, so the subdomains in each processor
contain the full radial extent of the grid. Our experiments confirmed
that using this domain decomposition, 
the number of iterations indeed does not depend much on the number 
of processors. Our largest test so far involves a $450\times 540\times 1200$
grid with 30 times more cells than the $150\times 180\times 360$ grids
discussed in most of this paper.
For the large problem we need about 8,500 iterations to reach the 
$10^{-10}$ relative accuracy, a factor of 9 increase relative to the smaller
problem. Using 108 CPU-s, the solution is obtained in about 5.3 hours.

Despite the various limitations, for some applications the spherical 
harmonics approach may still be preferred. For example if the solution is 
needed to obtain a spherical power spectrum of the solar magnetic field. 
If the solution is to be used in a finite difference code, 
the finite difference solution is probably preferable.  
We are using the FDIPS code to generate the potential field solution 
as the initial field for our solar corona model \citep{vanderHolst:2010}.

This paper attempts to call the attention of astrophysicists and solar
physicists to the limitations and potential pitfalls of using the spherical 
harmonics approach to obtain a potential field solution.
The spherical harmonics representation of the potential field solutions 
are available from several synoptic magnetogram providers, 
although the details of the method used to obtain the spherical harmonics 
is not always clear. 
A spherical harmonics based PFSS package implemented in IDL is available 
as part of the Solar-Soft library ({\tt http://www.lmsal.com/solarsoft}).
This package uses the magnetogram remeshing technique either onto 
the Chebyshev (uniform-$\theta$) or the Legendre collocation points.

We are not aware of any publically available code 
that uses finite differences to solve this particular
problem. To allow other researchers to use and compare the two approaches, 
we make our finite difference code FDIPS publically available at
the {\tt http://csem.engin.umich.edu/fdips/} website.

\clearpage

\end{document}